\documentclass[twocolumn,prb,superscriptaddress,showpacs,amsmath,amssymb,floatfix,aps]{revtex4}
\usepackage[english]{babel}
\usepackage[latin1]{inputenc}
\usepackage{graphics,graphicx,epsfig}% Include figure files
\usepackage{dcolumn}% Align table columns on decimal point

\newcommand {\rn} {$R$NiO$_3$}

\newcommand {\tmi} {T$_{MI}$}

\newcommand {\soni} {$\sigma ^2 _{Ni-O}$}
\newcommand {\sni} {$\sigma ^2 _{NiNi}$}

\begin{document}

\title{Short-range charge-order in $R$NiO$_{3}$ perovskites
($R$=Pr,Nd,Eu) probed by X-ray absorption spectroscopy}

\author{C\'{\i}nthia Piamonteze}
\affiliation{Laboratório Nacional Luz Síncrotron, Caixa Postal
6192 - 13084-971 - Campinas/SP, Brazil} \affiliation{IFGW/UNICAMP,
13083-970, Campinas/SP, Brazil}

\author{H\'{e}lio C. N. Tolentino}
\affiliation{Laboratório Nacional Luz Síncrotron, Caixa Postal
6192 - 13084-971 - Campinas/SP, Brazil}

\author{Aline Y. Ramos}
\affiliation{Laboratório Nacional Luz Síncrotron, Caixa Postal
6192 - 13084-971 - Campinas/SP, Brazil} \affiliation{LMCP-UMR 7590
- CNRS, Paris, France}

\author{N. E. Massa}
\affiliation{LANAIS, CEQUINOR, UNLP, C.C.962,
 1900 La Plata, Argentina}

\author{J.  A.  Alonso}
\author{M.  J. Mart\'{\i}nez-Lope}
\author{M.  T.  Casais}
\affiliation{Instituto de Ciencia de Materiales de Madrid,
C.S.I.C, Cantoblanco, E-28049 Madrid, Spain}

\date{\today}

\begin{abstract}

The short-range organization around Ni atoms in orthorhombic
$R$NiO$_{3}$ ($R$=Pr,Nd,Eu) perovskites has been studied over a
wide temperature range by Ni K-edge x-ray absorption spectroscopy.
Our results demonstrate that two different Ni sites, with
different average Ni-O bond lengths, coexist in those orthorhombic
compounds and that important modifications in the Ni nearest
neighbors environment take place across the metal-insulator
transition. We report evidences for the existence of short-range
charge-order in the insulating state, as found in the monoclinic
compounds. Moreover, our results suggest that the two different Ni
sites coexists even in the metallic state. The coexistence of two
different Ni sites, independently on the $R$ ion, provides a
common ground to describe these compounds and shed new light in
the understanding of the phonon-assisted conduction mechanism and
unusual antiferromagnetism present in all $R$NiO$_{3}$ compounds.

\end{abstract}


\maketitle

Lanthanide nickel perovskites ($R$NiO$_{3}$, $R$=lanthanide)
display, except for $R$=La, a thermally driven first-order
metal-insulator (MI) transition
\cite{Lacorre1991,Munoz1992,Torrance1992} and an unusual
antiferromagnetic order \cite{Munoz1994,Carvajal1998,Alonso1999}.
The detailed atomic and electronic structures, and their
modifications at the crossover from localized to itinerant
electronic behavior, remain among the most important questions to
be addressed. The crossover temperature (\tmi) to the metallic
state increases as the lanthanide ion decreases, showing a close
relation between \tmi\ and the structural distortion. These \rn\
compounds are at the boundary between low-$\Delta$ metals and
charge-transfer insulators \cite{Mizokawa1995,PiamontezeSRL}. The
charge-transfer energy $\Delta$ is smaller than the $3d-3d$
Coulomb repulsion, so that the band gap is controlled by
$\sim\Delta-W$, where $W$ is the one-electron bandwidth. Ni$^{3+}$
in low-spin configuration $(t_{2g}^{6}e_{g}^{1})$  is a
Jahn-Teller (JT) ion with a single $e_{g}$ electron with orbital
degeneracy. Owing to strong hybridization among the $Ni3d:e_{g}$
and $O2p:\sigma$ bands, some holes are transferred from the $3d$
to the $2p$ orbitals in the ground state, leading to partial
screening of the JT distortion.

The long-range structure of the lighter lanthanide
($R$=Pr,Nd,Sm,Eu) compounds is orthorhombic ($Pbnm$ space group),
with rather regular NiO$_{6}$ octahedra, in both metallic and
insulating states \cite{Munoz1992}. Upon decreasing the
temperature through the transition, a sudden slight increase in
the average Ni-O bond length
$(\bigtriangleup_{Ni-O}=0.004\textrm{Å})$, followed by the steric
accommodation in the tilts of the octahedra
$(\bigtriangleup\phi=-0.5^{o})$ was reported \cite{Munoz1992}. It
was proposed \cite{Torrance1992} that the closing of the gap stems
from the increase of the bandwidth $W$, which depends on the
Ni-O-Ni bond angle $(180^{o}-\phi)$ through $W\propto cos\phi$.
This description is incomplete since, in the metallic state,
infrared studies point to a conduction mediated by strong
electron-phonon coupling associated to local charge fluctuations
\cite{Massa1997b}. The significant shift observed in \tmi\ by
$^{16}$O-$^{18}$O isotope substitution \cite{Medarde1998} is an
additional indication of a JT-polaron assisted conduction
mechanism. Furthermore, our recent reports on the decreased
hybridization of the Ni-O bonding \cite{PiamontezeSRL} and on the
splitting in the coordination shell \cite{PiamontezePhysB}, in the
insulating state, emphasize that local modifications within the
NiO$_{6}$ octahedra play a fundamental role in the understanding
of the MI transition.

In the insulating state of the heavier lanthanide
($R$=Ho,Y,Er,Tm,Yb,Lu) compounds, the long-range structure is
monoclinic ($P2_{1}/n$ space group) \cite{Alonso1999,Alonso2000}.
The $P2_{1}/n$ symmetry implies the existence of small
Ni$^{\textrm{I}}$ and large Ni$^{\textrm{II}}$ sites. Alonso {\it
et al.} associated this monoclinic distortion to a long-range
charge-order along the three axes. The small site
(Ni$^{\textrm{I}}$) has stronger hybridization among $Ni3d$ and
$O2p$ orbitals than the large one (Ni$^{\textrm{II}}$). Moreover,
the large site (Ni$^{\textrm{II}}$) presents a larger JT
distortion, with a more localized single $e_{g}$ electron. Zhou
and Goodenough \cite{Zhou2004} suggested that the difference in
charge-transfer between the $3d$ and $2p$ orbitals in each Ni site
would give rise to the charge-order. The existence of
charge-ordered Ni sites establishes a favorable scenario to
understand the unusual magnetic structure that requires
alternating couplings \cite{Munoz1994,Carvajal1998,Alonso1999}.
Alonso {\it et al.} found different magnetic moments for each Ni
site: 0.7 and 1.4$\mu_B$ for the small and large Ni sites in
YNiO$_3$, respectively \cite{Alonso1999}.

Concerning lighter lanthanide compounds, attempts to observe
long-range charge-order by neutron diffraction have been
unsuccessful \cite{Munoz1994,Carvajal1998,Medarde1998}. Recently,
electron diffraction and Raman scattering \cite{Zaghiroui2001} and
resonant x-ray scattering \cite{Staub2002} studies produced
evidences on a small monoclinic distortion in NdNiO$_{3}$ thin
films. Even if the extension of these results to polycrystalline
compounds is not straightforward, they question the validity of
the assignment of the orthorhombic symmetry and single magnetic
moment associated to Ni in the lighter lanthanide compounds. More
recently, evidences that the low temperature distortion is shared
by all members of the \rn\ family have been presented
\cite{delaCruz2002}.

We report on an EXAFS (Extended X-Ray Absorption Fine Structure)
spectroscopy study on the lighter lanthanide ($R$=Pr,Nd,Eu)
compounds. We present evidences of the coexistence of two
different Ni sites in these orthorhombic compounds, over a wide
temperature range, providing a common background to describe the
whole \rn\ family. The polycrystalline samples studied were
described elsewhere \cite{Alonso1995}. EXAFS measurements were
carried out at the XAS beam line of LNLS, Brazil
\cite{Tolentino2001}, using a Si(111) monochromator, with an
energy resolution of 2.4eV at Ni K edge. Spectra were measured in
transmission mode up to a maximum wavenumber of
$k\simeq19$\AA$^{-1}$. A cryostat-heater was used to control the
temperature from 10K to 600K.

\begin{figure}
\begin{center}
\epsfig{file= 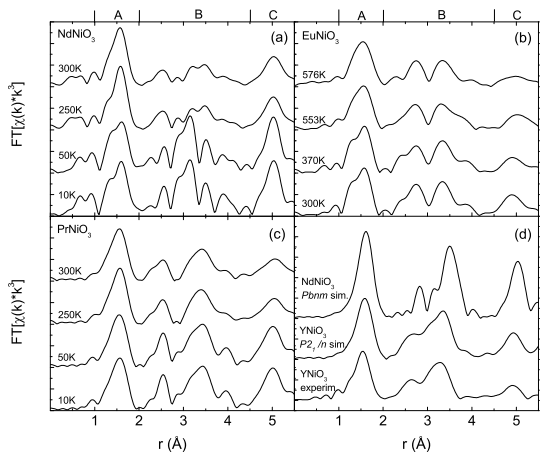,width=8.0cm} \caption{\label{fig1}
Fourier Transform of Ni K-edge EXAFS signal for (a) NdNiO$_{3}$,
(b) EuNiO$_3$, (c) PrNiO$_3$ at selected temperatures and for (d)
YNiO$_3$ at room temperature. Simulations for $P2_{1}/n$ and
$Pbnm$ space groups also shown in (d).}
\end{center}
\end{figure}

\begin{figure}
\begin{center}
\epsfig{file=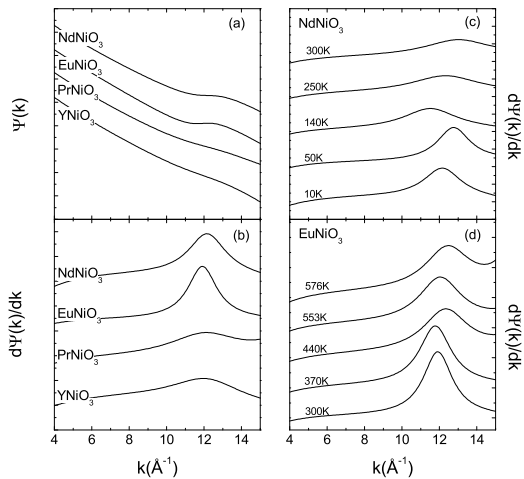,width=7.0cm} \caption{\label{fig2}(a)
Phase and (b) phase derivative (PD) functions for $R$NiO$_{3}$ in
the insulating state. PD functions for (c) NdNiO$_{3}$ and (d)
EuNiO$_{3}$ across the MI transition.}
\end{center}
\end{figure}

Figure \ref{fig1} shows the Fourier Transform (FT) amplitude of
the experimental EXAFS signal for NdNiO$_{3}$ ($T_{MI}=200K$),
EuNiO$_3$ ($T_{MI}=460K$) and PrNiO$_3$ ($T_{MI}=130K$) at a few
selected temperatures, with two spectra in the metallic and two in
the insulating states. The FT represents a pseudo-radial
distribution function around Ni atoms, where the r-values are
shifted by a small amount due to the phase-shift of the
photoelectron wave function \cite{Rehr2000}. A clear splitting,
associated to the existence of separable bond lengths in the
coordination shell (peak A around 1.6$\textrm{Å}$), is observed
for both NdNiO$_{3}$ and EuNiO$_{3}$ in the insulating state
(fig.\ref{fig1}-a,b). The splitting is not well resolved but an
asymmetry in the low-$r$ values is observed for PrNiO$_{3}$
(fig.\ref{fig1}-c). In the metallic state (fig.\ref{fig1}-a,b,c),
the peak A is asymmetric for all compounds. We also show the FT
amplitude for the monoclinic insulating compound YNiO$_{3}$
($T_{MI}=582K$) and two simulations using Feff7 \cite{Rehr2000}
(fig.\ref{fig1}-d). The peak A for YNiO$_{3}$ displays a similar
asymmetry. Simulations have been performed based on the
crystallographic structure found by neutron diffraction for
YNiO$_{3}$ \cite{Alonso1999} and NdNiO$_{3}$ \cite{Munoz1992}. The
simulation for YNiO$_{3}$ with the $P2_{1}/n$ space group, which
includes the two non-equivalent Ni sites, reproduces very well the
overall YNiO$_{3}$ spectrum (fig.\ref{fig1}-d), with a slight
asymmetry in the coordination shell. On the other hand, the
simulation for NdNiO$_{3}$ with the $Pbnm$ space group
(fig.\ref{fig1}-d) produces a sharper coordination shell, with no
asymmetry nor splitting at all, as expected from a single Ni
environment. The comparison of the NdNiO$_{3}$($Pbnm$) simulation
with the experimental low-temperature NdNiO$_{3}$ spectrum
(fig.\ref{fig1}-a) demonstrates that a single Ni site cannot
describe the short-range Ni environment. On the other hand, the
medium and long-range structure is quite well reproduced with the
orthorhombic symmetry, showing that the splitting is really taking
place only at the short-range scale.

The EXAFS signal obtained from the combination of two Ni sites,
with identical backscattering atoms but with slightly different
average bond lengths, presents a beating in the amplitude function
\cite{Martens1977}. The wavenumber where the beating occurs
($k_{b}$) satisfies $\delta r.k_{b}\approx\pi/2$, where $\delta r$
is the separation in bond length \cite{Martens1977,Jaffres2000}.
This beating in the EXAFS amplitude generates a splitting in the
FT amplitude (as seen in peak A, fig.\ref{fig1}). The phase
function $\Psi(k)$, extracted from the peak A analysis by the
inverse-FT, gives an inflexion point that leads to an extremum in
the phase derivative (PD) function (fig.\ref{fig2}-a,b). The PD
analysis is a very sensitive tool to follow small separations
between two close distances, particularly in distorted perovskites
\cite{Souza2004}. Figure \ref{fig2}-a shows the phase functions
extracted by back-transforming the peak A (fig.\ref{fig1}) taking
the average Ni-O bond lengths for the $R$NiO$_{3}$
($R$=Nd,Eu,Pr,Y) compounds. Maxima appear around the beating
position $k_{b}\simeq12.3$\AA$^{-1}$ in the PD function
(fig.\ref{fig2}-b) and give an unambiguous signature of the
existence of the two well separated Ni-O bond lengths. An
estimation of the separation value of roughly $\delta
r\approx0.12\textrm{Å}$ can be deduced from the beating position.
This value is affected by systematic errors \cite{Jaffres2000} and
is slightly overestimated. The average separation found for
YNiO$_{3}$ and for all heavier lanthanide compounds
\cite{Alonso1999,Alonso2000} is $\delta r\approx0.09\textrm{Å}$.
This value can be used to correct for systematic errors and
calibrate the scale. What is important to remark here is that the
PD maximum occurs at about the same position for all compounds.

The PD analysis, as well as the FT splitting, supports the
coexistence of two Ni sites with different average bond lengths.
The lighter lanthanide ($R$=Pr,Nd,Eu) compounds, as the heavier
lanthanide ones, present small and large Ni sites with a similar
separation of about $\delta r\approx0.09\textrm{Å}$ among them.
Owing to the similarity among all \rn\ compounds, we conclude
that, at the short-range scale probed by EXAFS, a similar
charge-order modulation takes place in the insulating state.
Nevertheless, only in the heavier lanthanide compounds this
charge-order modulation develops at long-range and gives rise to
the lowering of the crystallographic symmetry from orthorhombic to
monoclinic. The behavior of the PD function with increasing
temperature for NdNiO$_{3}$ and EuNiO$_{3}$ (fig.\ref{fig2}-c,d)
shows that the amplitude at the inflexion point decreases but does
not vanish and remains around the same position for $T>T_{MI}$.
Above the first-order transition, in the metallic state,
conduction electrons moving through the lattice destroy
charge-order and the amount of split Ni sites diminishes. But, two
different Ni bond lengths, with roughly the same separation, are
still present. This result suggests that, in the metallic state,
regions of split Ni sites survive and coexist immersed in the
conducting matrix, in accordance with a recent infrared
spectroscopy study \cite{delaCruz2002}. The coexistence of more
localized electrons in a Fermi-liquid background have also been
suggested to be a common feature in \rn\ compounds
\cite{Zhou2000prb,Zhou2003}. These phase inhomogeneities should be
delocalized and fluctuating within the material. The conduction
mechanism, which is strongly coupled to the lattice
\cite{Medarde1998}, is intimately related to these fluctuations.

\begin{figure}
\begin{center}
\epsfig{file=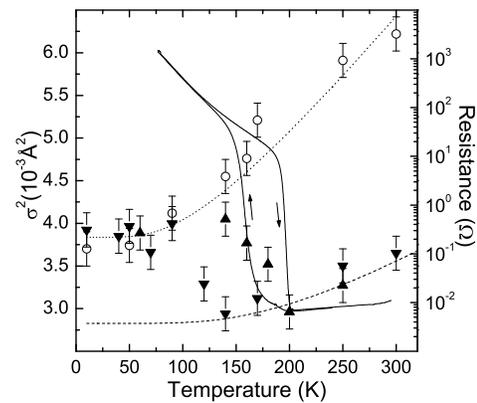,width=7.0cm}
\caption{\label{fig3}Total disorder for the NdNiO$_{3}$
coordination shell while decreasing ($\blacktriangledown$) and
increasing ($\blacktriangle$) the temperature, and for Ni shell at
5$\textrm{Å}$ ($\circ$). Thermal behavior for the coordination
shell ($---$) and Ni shell ($...$), and resistance (---).}
\end{center}
\end{figure}

The structural crossover can be studied by the Debye-Waller factor
$exp(-2\sigma^{2}k^{2})$, which damps the EXAFS signal and the FT
amplitude. This factor contains static and thermal contributions
\cite{Rehr2000}. The static contribution comes from the bond
length dispersion around the average value. The thermal
contribution is related to the lattice vibrations, which decreases
with temperature and can be well described by correlated Einstein
or Debye models \cite{Sevillano1979}. To quantify the
modifications observed in the FT amplitude (fig.\ref{fig1}), we
performed an EXAFS analysis limited to $k=12$\AA$^{-1}$ using
theoretical amplitude and phase functions \cite{Rehr2000} and an
average Ni coordination shell. This restricted k-range
characterizes a low-resolution study in $r$-space, where the Ni-O
bond lengths are no longer distinguishable and the bond length
separation appears as a static contribution to the total disorder.
Figure \ref{fig3} shows the total disorder of the coordination
shell (\soni, peak A in fig.\ref{fig1}) and of the fifth Ni shell
(\sni, peak C in fig.\ref{fig1}) for NdNiO$_{3}$. \soni\ follows
the expected thermal behavior from room temperature down to 170K,
then smoothly increases down to 90K and remains constant. When the
temperature is increased, a structural hysteresis is observed,
consistent with the first-order character of the transition,
illustrated by the measured resistance with temperature.
Subtracting the extrapolated thermal behavior from the total
disorder of the coordination shell, one gets the increase in bond
length dispersion from the metallic to the insulating state. This
experimental increase is about 0.0012\AA$^{2}$ (fig.\ref{fig3}).
This value is related to the amount of split Ni sites changing at
the crossover. If one considers the crossover from a "pure
metallic" state (one single Ni site with bond length of 1.95\AA)
to a "pure insulating" state (long-range charge-ordered
Ni$^{\textrm{I}}$ and Ni$^{\textrm{II}}$ sites with bond length
separation of 0.09\AA) an increase in bond length dispersion of
$\bigtriangleup$\soni $\approx0.0020$\AA$^{2}$ should be found.
This is much larger than the experimental value (0.0012\AA$^{2}$),
indicating that only a fraction of Ni sites is splitting at the
crossover. On the other hand, \sni\ exhibits the expected smooth
thermal decrease with temperature, showing that the splitting in
bond length takes place only at the short-range scale restricted
to the coordination shell. Similar behaviors have been obtained
for EuNiO$_{3}$ and PrNiO$_{3}$.

The local structural modification evidenced by \soni\ takes place
in a broader temperature range (80K) than that of resistance
measurement (25K). The sharp electronic transition is then related
to a smoother local structural modification. Across the
first-order MI transition, the system gets into a state where the
volume fraction of the localized-electron phase increases but the
metallic phase does not disappear abruptly. The coexistence of
metallic and non-metallic phases has already been invoked to
explain some peculiarities of the electrical resistivity and
Seebeck coefficient in NdNiO$_{3}$ \cite{Granados1993}. In \rn\
compounds, we believe that regions of a localized-electron phase
exist in the metallic matrix above \tmi, but the conduction
electrons prevent the charge-order state and the split sites are
fluctuating. When the temperature goes below \tmi, the electron
localization triggers the transition and the volume fraction of
these regions grows. The system undergo a structural phase
transition into a long-range charge-ordered state, with $P2_{1}/n$
symmetry, in the case for the heavier lanthanide compounds
\cite{Alonso1999,Alonso2000}. In the case of the lighter
lanthanides, the modulation is limited to a short-range scale. The
compound remains disordered at long-range and there is no breaking
in the $Pbnm$ symmetry \cite{Munoz1992}. The presence of
JT-distorted sites and short-range charge-order in the insulating
state of the orthorhombic compounds gives an explanation to the
strange antiferromagnetic arrangement that requires two
nonequivalent Ni sites \cite{Munoz1994,Carvajal1998}. As in the
monoclinic compounds, two different magnetic moments can now be
associated to Ni. In both orthorhombic and monoclinic compounds,
the crossover from the insulating to the metallic state can be
regarded as a melting of the charge-order but with important local
dynamical distortion fluctuations still remaining. The persistence
of these fluctuations, associated to local distortion of NiO$_{6}$
octahedra, explains the phonon-assisted conduction mechanism in
the metallic state, as suggested by Medarde {\it et al.}
\cite{Medarde1998}.

In conclusion, we demonstrated by using EXAFS spectroscopy that
important modifications take place in the Ni coordination shell of
\rn\ compounds across the MI transition. Our results give
evidences of the coexistence of two different Ni sites and
short-range charge-order in the insulating states of
polycrystalline $R$NiO$_{3}$ (R=Pr,Nd,Eu) orthorhombic compounds,
as earlier found in the monoclinic systems. The existence of
short-range charge-order in these compounds provides a common
ground to describe their transport properties and unusual
antiferromagnetism. The coexistence of two phases in the metallic
state, a fluctuating localized-electron phase with JT-distorted
sites immersed in a conducting matrix, shed new light in some
intriguing properties, like the phonon-assisted conduction
mechanism.

Work partially supported by LNLS/ABTLuS. CP thanks FAPESP for her
PhD grant.

%\bibliography{C:/Arq/Helio/Artigos_preparation/PRB_RapidComm_NiK/RNOc}

\end{document}